\documentclass{elsart}
\usepackage{epsfig}
\journal{Journal of non-crystalline solids}

\begin{document}
\begin{frontmatter}

\title{Dynamical heterogeneities in an attraction driven colloidal glass}
\author[antonio]{Antonio M. Puertas}\ead{apuertas@ual.es}, 
\author[matthias]{Matthias Fuchs}, 
\author[mike]{Michael E. Cates}

\address[antonio]{Group of Complex fluids physics, Department of Applied Physics,
University of Almer\'{\i}a, 04120 Almer\'{\i}a, Spain}
\address[matthias]{Fachbereich Physik, University of Konstanz, D-78457 Konstanz, 
Germany}
\address[mike]{SUPA, School of Physics, The University of
Edinburgh, JCMB Kings Buildings, Mayfield Road, Edinburgh EH9 3JZ, UK}
\date{\today}

\begin{abstract}
The dynamical heterogeneities (DH) in non-ergodic states of an attractive 
colloidal glass are studied, as a function of the waiting time. 
Whereas the fluid states close to vitrification showed strong DH, the distribution 
of squared displacements of the glassy states studied here only present a tail 
of particles with increased mobility for the lower attraction strength at short
waiting times. These particles are in the surface of the percolating cluster that 
comprises all of the particles, reminiscent of the fastest particles in the fluid. 
The quench deeper into the attractive glass is dynamically more 
homogeneous, in agreement with repulsive glasses (i.e. Lennard-Jones glass).
\end{abstract}


\end{frontmatter}

\section{Introduction}

Two distinct glasses have been predicted and identified in hard spheres
with short range attractions: a repulsion driven glass at high density, and
an attraction driven one, formed at low temperatures (or high attraction
strength) \cite{pham02,sciortino02}. Whereas the former one is caused by the steric
hindrance of the particle cores and the so-called {\sl cage effect}, the latter 
forms due to the bonding between particles. This system is realized experimentally
by a colloid-polymer mixture, where the effect of the polymers is to induce
an effective attraction between the colloids \cite{likos}. Both glasses have
been indeed identified, although the attractive one, which at low concentrations is termed `gel', often 
competes with (and inhibits) liquid-gas phase separation \cite{pham04}.

Dynamical heterogeneities (DH) have been found in the proximity of repulsion
driven glass transitions by computer simulations, i.e. in Lennard-Jones
mixtures \cite{kob97,flenner05}, or hard spheres \cite{doliwa98}. In these cases, 
while the system is structurally liquid-like (homogeneous), a population of
particles of increased mobility is observed. As the glass transition is
approached from the fluid side, the heterogeneities become more pronounced,
but decrease again deeper in the glass \cite{weeks00,vollmayr02,vollmayr05}. The role of these
dynamical heterogeneities in the glass transition is as yet unclear; whereas
mode coupling theory focusses on averaged quantities and neglects them \cite{gotze91}, the so-called
facilitated dynamics theories give DH the central role for their description
of the glass transition \cite{berthier03}.

In recent works, it has been shown that DH can be found also in attractive 
glasses, by studying the distribution of particle displacements in the system 
\cite{puertas04,puertas05,reichman05}. In fluid states close to the transition 
two populations of particles were found, separated by a minimum in the 
displacement distribution. A similar feature has been found also in
repulsive glasses, which could imply a common origin \cite{flenner05,reichman05}. 
However, the low density of the attractive glass, as low as $\phi_c=0.4$, causes structural 
heterogeneities as well; the system forms a percolating cluster of high density 
material, leaving voids with no particles. A correlation between
structural and dynamical heterogeneities is thus possible, showing that
`fast' particles are in the surface of the cluster, whereas the `slow' ones
are mostly trapped in the inner parts of it \cite{puertas04}.

In this work, we study the DH inside the non-ergodic
region, for two different states, and compare them with those of the
equilibrium systems. Only one population of particles can be identified from
the distribution of particle displacements, and the distribution is narrower
for the state with stronger attractions. Moreover, as the systems age, they
become more and more homogeneous, from the point of view of the dynamics.
Both results indicate that the strongest DH are
obtained in the fluid side of the phase diagram, close to the glass
transition. As a side remark, it must be noted that the structural
heterogeneities mentioned above persist in the out-of-equilibrium systems,
and thus are not the sole origin of the DH in attractive glasses.

\section{Simulation details}

We have performed computer simulations of a system composed of $1000$ soft 
core ($V_C\sim r^{-36}$) particles with attractive interactions given by the 
Asakura-Oosawa (AO) potential \cite{likos}. It models a mixture of colloids 
with non-adsorbing polymers, and the range of attraction is set by the polymer size. 
In order to guarantee full access to the whole parameter space, phase 
separations have been inhibited. Crystallization is avoided by polydispersity 
(flat distribution, $\delta=10\%$ width), and liquid-gas demixing by a  
repulsive barrier extending to two mean diameters. Further details of the 
interaction potential can be found in previous works \cite{puertas05}.

Length is measured in units of the average radius, $a$, and time in units of 
$\sqrt{4a^2/3v^2}$, where the thermal velocity, $v$, was set to $\sqrt{4/3}$. 
Equations of motion were integrated using the velocity-Verlet algorithm, in the 
canonical ensemble (constant NTV), to mimic the colloidal dynamics, with a time 
step equal to $0.0025$. Every $n_t=100$ time steps, the velocity of the particles was
re-scaled to assure constant temperature. The range of the attraction is set to 
$0.2a$. Density is reported as volume fraction, $\phi_c=4/3\pi a^3 
\left(1+\left(\delta/a\right)^2\right) n_c$, with $n_c$ the number 
density, and the attraction strength is measured in units of the polymer volume
fraction $\phi_p$ (at contact the AO attraction strength is $16 k_BT \phi_p$).
The attractive glass transition for this system has been studied previously 
\cite{puertas05,puertas05b}. An MCT analysis of the results (diffusion 
coefficient, time scale and viscosity) yields a transition point at 
$\phi_p=0.4265$ for the colloid density $\phi_c=0.40$.

For the study of aging here, the systems were equilibrated without attraction 
($\phi_p=0$) at $\phi_c=0.40$, and then 
instantaneously {\sl quenched} to the desired $\phi_p$ at zero time, $t=0$. Two
attraction strengths have been studied, $\phi_p=0.50$ and $\phi_p=0.80$, lying 
beyond the nonergodicity transition. In both cases, $25$ independent simulations 
have been 
performed, and the evolution of the system has been followed as a function of 
the time elapsed since the quench, called waiting time, $t_w$. Correlation 
functions thus depend on two times: $t_{w}$ and $t'=t-t_{w}$.

\section{Results and discussion}

\begin{figure}
\begin{center}
\includegraphics*[width=13cm]{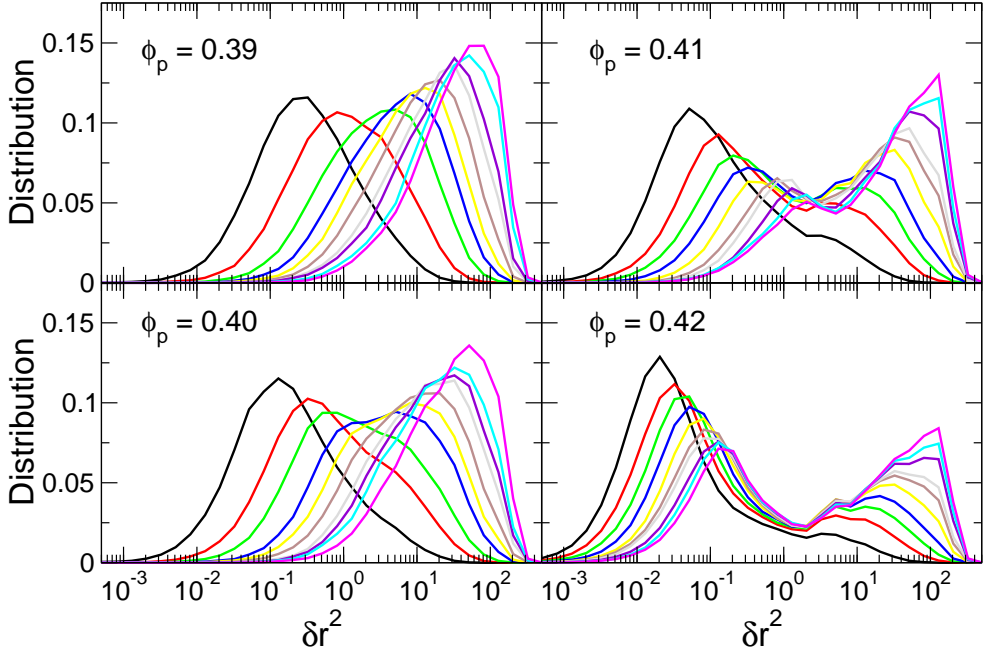}
\end{center}
\caption{\small Distribution of squared displacements at different times: from left
to right, $t=100$, $200$, $400$, $800$, ..., and for different polymer
fractions, $\phi_p$, as labeled. Note that as $\phi_p$ increases, two
populations of particles with different mobilities appear in the system. The
glass transition is located at $\phi_p=0.4265$, estimated from MCT analysis
(power law fittings) \cite{puertas05}.}
\label{fig1}
\end{figure}

In fluid states close to the attractive glass, increasing DH have been found, 
the stronger the attraction \cite{puertas04}. Two populations of particles are 
observed to appear as the attraction is increased, one of mobile particles and 
another one of quasi-immobile particles (see Fig. \ref{fig1}) \cite{puertas05}. 
(The minimum in the distribution of squared displacements allows for an 
unambiguous identification of almost every particle.)
These two populations are structurally well differentiated: the particles in
the `skin' of the gel, with a small number of bonds, are mobile, and the
particles in the inner parts of the gel, the `skeleton', are quasi-immobile.
The populations were observed to be stable for long times, although
particles can change from one to the other. It is thus an equilibrium feature
of the system, partly induced by the structural heterogeneity. 

\begin{figure}
\begin{center}
\includegraphics*[width=13cm]{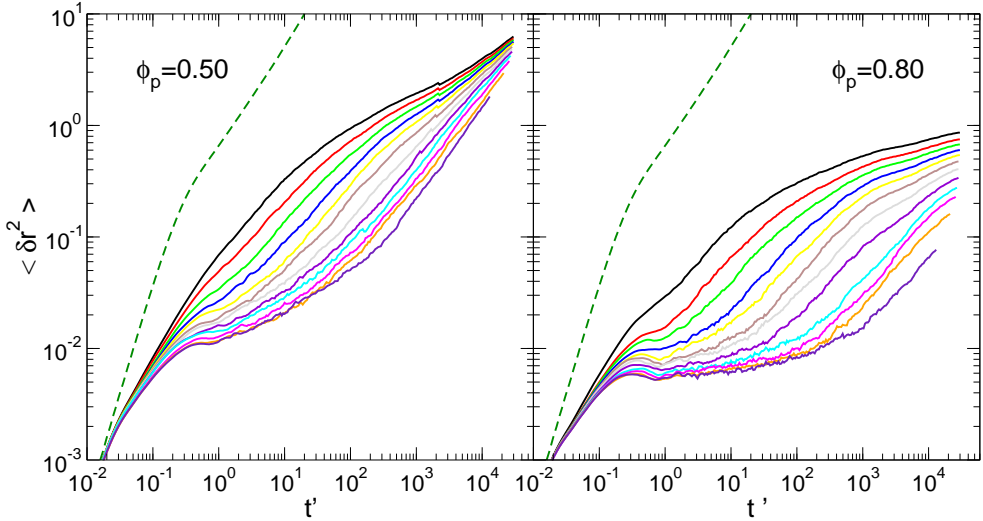}
\end{center}
\caption{Mean squared displacements for $\phi_p=0.50$ and $\phi_p=0.80$, as labeled, 
for different waiting times; from left to right, $t_w=8$, $16$, $32$, $64$, $\dots$, 
$16384$. The dashed line shows the MSD for hard spheres at this density.}
\label{msd}
\end{figure}

\begin{figure}
\begin{center}
\includegraphics*[width=10cm]{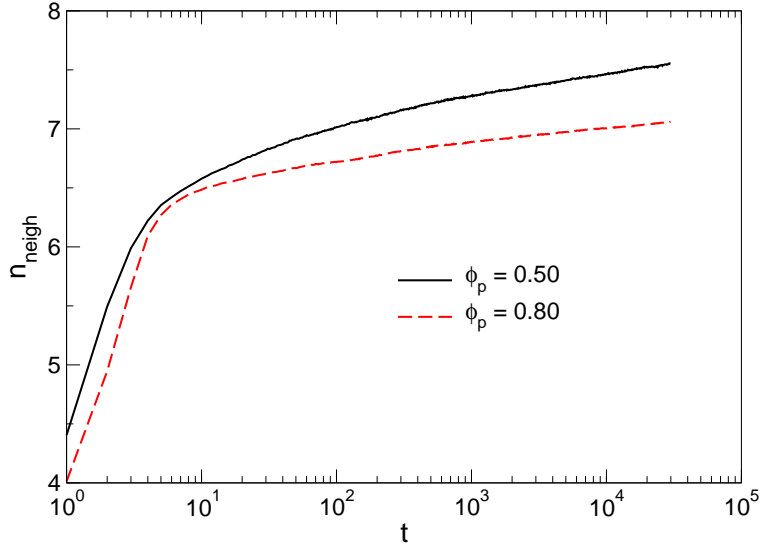}
\end{center}
\caption{Evolution of the mean number of neighbours per particle in the system after the quench for both states, as labeled.}
\label{nneigh}
\end{figure}

\begin{figure}
\begin{center}
\includegraphics*[width=11cm]{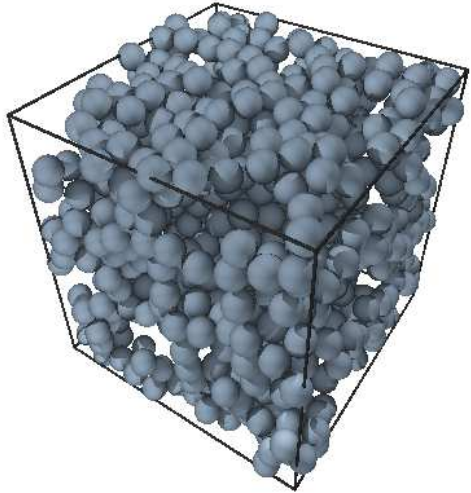}
\end{center}
\caption{Snapshot of the system with $\phi_p=0.50$ at a long time, $t=30000$. 
(Qualitatively similar snapshots are obtained for $\phi_p=0.80$.)}
\label{gel}
\end{figure}

All of the states presented in Fig. \ref{fig1} are, nevertheless, fluid states.
The structural properties do not depend on time, nor the dynamical ones on the
initial time, and the correlation functions averaged throughout the whole
system decay to zero. By further increasing the attraction strength, however,
the system falls out of equilibrium and shows aging. In Fig. \ref{msd}, the mean 
squared displacement (MSD) is presented for the two quenches studied here, 
$\phi_p=0.50$ and $\phi_p=0.80$; different lines are the MSD for different
waiting times, $t_w$. As $t_w$ increases, bonds are formed between particles, as shown in Fig. \ref{nneigh} for both states, which hinders the motion of the particles, causing the dynamical arrest.
 
In the MSD, a plateau develops at short distances, signaling the localization 
length, clearer in the case of the higher $\phi_{p}$, where the bonds are stronger and the {\sl cage} (network of bonds) of the particles is tighter, as shown by the shorter localization length 
(implying higher non-ergodicity parameters in the density correlation function). The mean number of bonds per particle is, however, lower. Also, the aging is more dramatic and at long waiting times the MSD hardly reaches 
the range of the attraction, contrary to previous works where the plateau in MSD, or
the density correlation function, was not observed \cite{zaccarelli03}.
Here we will concentrate on the DH, and a full 
analysis of the aging will be presented elsewhere \cite{puertas06}.

\begin{figure}
\begin{center}
\includegraphics*[width=11cm]{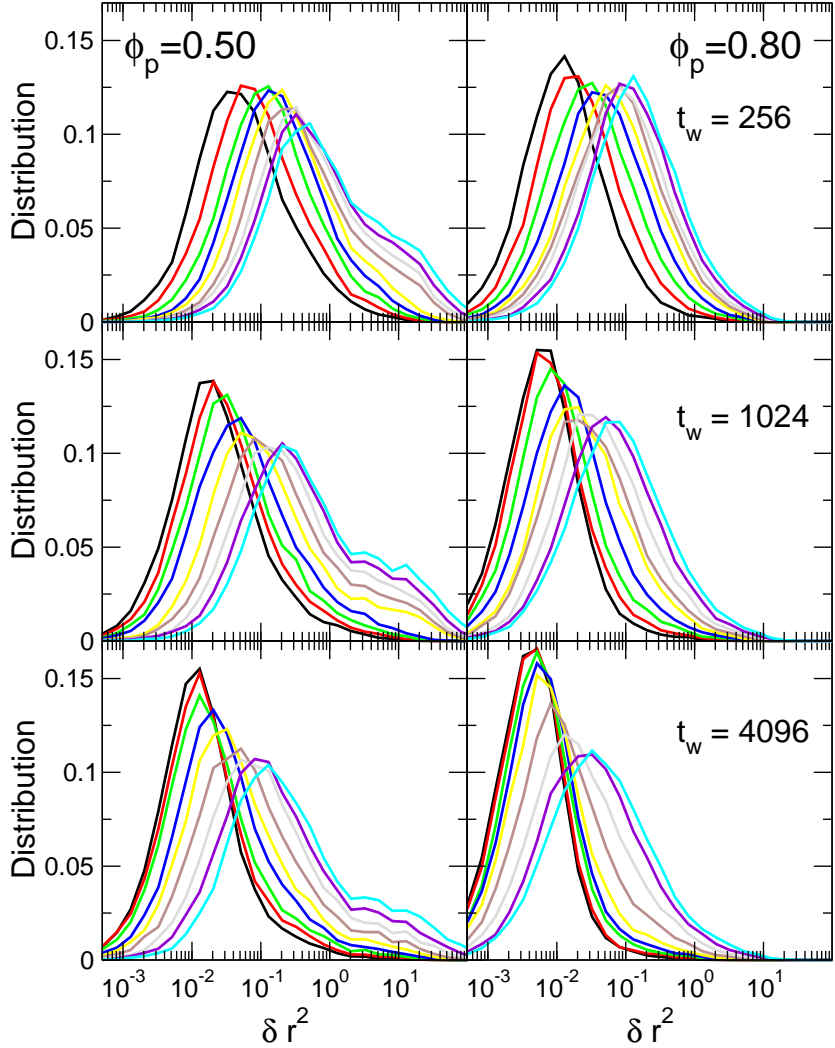}
\end{center}
\caption{Distribution of squared displacements for $\phi_p=0.50$ and $\phi_p=0.80$ 
and three waiting times, as labeled. The different lines represent the distribution 
at $t=100$, $200$, $400$, $800$, $1600$, $3200$, $6400$, $12800$ and $25600$, from 
left to right, respectively.}
\label{distributions}
\end{figure}

\begin{figure}
\begin{center}
\includegraphics*[width=13cm]{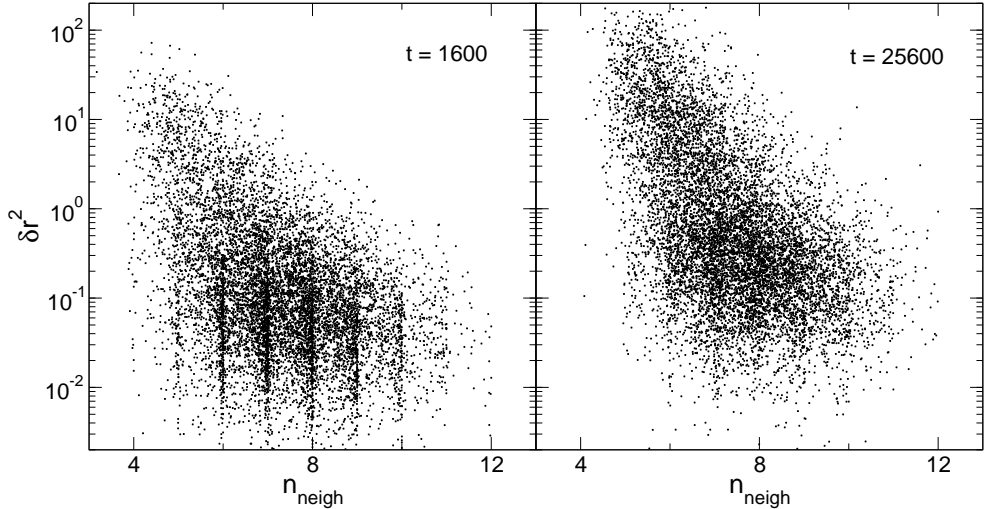}
\end{center}
\caption{Squared displacements the particles during time $t$, as a function of its
mean number of neighbours during this time, for $\phi_p=0.50$ and $t_w=1024$. 
(Only ten simulations are considered, i.e. $10000$ particles). Note the vertical bands 
at integer numbers of neighbours (especially for $t=1600$), due to particles that have not
changed their neighbours.}
\label{nneigh-msd}
\end{figure}

A snapshot of the system with $\phi_p=0.50$ is presented in Fig. \ref{gel}. The 
system forms an intricate structure, with voids and tunnels, similar to the fluid 
states presented above. Therefore, the ``skeleton and skin" picture presented above, 
with two different populations of particles, could still be applicable. In Fig. 
\ref{distributions} we present the distribution of squared displacements at 
different times for both quenches. Strikingly, the distribution is monomodal in
all cases, 
and the peak evolves to larger displacements as time proceeds. The strong DH 
observed in the fluid states, have, therefore, disappeared in the glass side. 
Moreover, the deeper quench, $\phi_p=0.80$, shows narrower distributions at 
all waiting times than the quench at $\phi_p=0.50$, indicating that the dynamics is 
more homogeneous the deeper the state is in the glass, in agreement with findings for Lennard-Jones 
(repulsive) glasses \cite{vollmayr05}. 

At $\phi_p=0.50$, however, a tail in the distribution to long distances can be 
observed, caused by some particles that can travel long 
distances; the number of which decreases with $t_w$. This feature is reminiscent 
of the population of fast particles observed in the fluid (Fig. \ref{fig1}). The 
origin of this population is studied in Fig. \ref{nneigh-msd}, where the squared 
displacement of every particle for a given time is correlated with the mean number 
of neighbours of the particle during this time. The plot shows that indeed the 
fastest particles in the system have less neighbours on average, whereas the 
particles with many neighbours, move very little. Therefore, the simple picture 
of ``skeleton and skin", can still be applied for the attractive glass close to the
transition. The number of these fast particles is, nevertheless, decreasing with 
waiting time, as observed by comparing similar times $t$ for different waiting times
$t_w$ in Fig. \ref{distributions}. Accordingly, the mean number of bonds per particle increases (see Fig. \ref{nneigh}), implying a compaction in the system, and thus having fewer particles in the skin, and even those, more tightly trapped. However, we cannot state whether the population of fast particles will reach a steady state or if it will vanish eventually at very long waiting times. At $\phi_p=0.80$, on
the other hand, the tail of fast particles is absent in the distribution of squared
displacements, although similar structural heterogeneities are observed.

\section{Conclusions}

We have shown that the strong DH found in fluid states close to the attractive glass
transition in colloids with short range attractions decrease again deep into the
non-ergodic region. The distribution of squared displacements is monomodal and no
particles with increased mobility are observed ($\phi_p=0.80$). However, close to 
the glass transition, in the glass side, aging is slower and some DH can still be 
detected: a tail in the distribution of squared displacements indicates 
fast particles, that can be identified with particles in the outer parts of the 
particle network. This feature is reminiscent of the population of fast particles
found in the fluid states close to the transition. However, the stability of this
population of fast particles in the glass cannot be established, but should not be
present in a truly arrested glassy state. The results presented here agree with those 
of repulsive glasses, contrary to other comparions between attractive and repulsive 
glasses \cite{zaccarelli03}.

{\sc Acknowledgements}

A.M.P. acknowledges financial support by the DGCYT (project MAT2003-03051-CO3-01). 
This work was funded in part (M.E.C.) by EPSRC GR/S10377. A.M.P. and M.F. were partially
funded by AI-DAAD project no. HA2004-0022.

\end{document}